\documentclass[12pt,preprint]{aastex}
\usepackage{graphicx}
\usepackage{amssymb}

\shorttitle{Particle acceleration in Protostellar Jets}
 \shortauthors{Rodriguez-Kamenetzky et al.}

\begin{document}

\title{Investigating Particle Acceleration in Protostellar Jets: \\
      The Triple Radio Continuum Source in Serpens}

 \author{Adriana Rodr\'{\i}guez-Kamenetzky\altaffilmark{1,2}, Carlos Carrasco-Gonz\'alez\altaffilmark{2}, Anabella Araudo\altaffilmark{3}, Jos\'e M. Torrelles\altaffilmark{4 \dagger}, Guillem Anglada\altaffilmark{5}, Josep Mart\'{\i}\altaffilmark{6}, Luis F. Rodr\'{\i}guez\altaffilmark{2}, Carlos Valotto\altaffilmark{1}}

  \altaffiltext{1}{Instituto de Astronom\'ia Te\'orica y Experimental, (IATE-UNC), X5000BGR C\'ordoba, Argentina} 
  \altaffiltext{2}{Instituto de Radioastronom\'ia y Astrof\'isica (IRyA-UNAM), 58089 Morelia, M\'exico} 
  \altaffiltext{3}{University of Oxford, Astrophysics, Keble Road, Oxford OX1 3RH, UK}
  \altaffiltext{4}{Institut de Ci\`{e}ncies de l'Espai (CSIC-IEEC) and Institut de Ci\`{e}ncies del Cosmos (UB-IEEC), Mart\'{i} i Franqu\`{e}s 1, 08028 Barcelona, Spain}
  \altaffiltext{5}{Instituto de Astrof\'{\i}sica de Andaluc\'{\i}a, CSIC, Camino Bajo de Hu\'etor 50, E-18008 Granada, Spain}
  \altaffiltext{6}{Dept. de F\'{\i}sica, EPS de Ja\'en, Universidad de Ja\'en, Campus Las Lagunillas s/n, A3-402, 23071 Ja\'en, Spain}
  \altaffiltext{$\dagger$}{The ICC (UB) is a CSIC-Associated Unit through the ICE}

\begin{abstract}
While most protostellar jets present free-free emission at radio wavelengths, synchrotron emission has been also proposed to be present in a handful of these objects. The presence of non-thermal emission has been inferred by negative spectral indices at centimeter wavelengths. In one case (the HH 80-81 jet arising from a massive protostar), its synchrotron nature was confirmed by the detection of linearly polarized radio emission. One of the main consequences of these results is that synchrotron emission implies the presence of relativistic particles among the non-relativistic material of these jets. Therefore, an acceleration mechanism should be taking place. The most probable scenario is that particles are accelerated when the jets strongly impact against the dense envelope surrounding the protostar. Here, we present an analysis of radio observations obtained with the Very Large Array of the Triple Radio Source in the Serpens star-forming region. This object is known to be a radio jet arising from an intermediate-mass protostar. It is also one of the first protostellar jets where the presence of non-thermal emission was proposed. We analysed the dynamics of the jet as well as the nature of the emission and discuss these issues in the context of the physical parameters of the jet and the particle acceleration phenomenon.
\end{abstract}

\keywords{Particle acceleration--- ISM: jets and outflows - Star formation}

\section{Introduction}

 Jets from young stellar objects (YSOs) have been long studied at radio wavelengths (e.g., Rodr\'{\i}guez 1995, 1996; Anglada 1996; Carrasco-Gonz\'alez et al. 2012; Anglada et al. 2015). Free-free interactions between thermal electrons and protons in these partially ionized jets produce detectable free-free continuum emission at centimeter wavelengths with a characteristic positive spectral index ($\alpha$, defined as S$_{\nu}$ $\propto$ $\nu^{\alpha}$). High angular resolution observations at radio wavelengths allow the jet material to be traced up to a few tens of astronomical unit (AUs) from the protostar. In this way, the base of the large pc-scale jets (which cannot be observed at optical or IR wavelengths due to the large extinction near the protostar) can be studied. These radio sources are called thermal radio jets, in contrast to the radio jets usually observed emanating from active galactic nuclei (AGNs) whose main emission mechanism at radio wavelengths is of non-thermal nature (optically thin synchrotron emission) and show very different characteristics: negative spectral indices at cm wavelengths, and linear polarization. 

 In the last decades, radio emission with negative spectral indices at centimeter wavelengths has also been detected in several YSO jets, such as the Triple Radio Source in Serpens (Rodr\'{\i}guez et al. 1989), HH 80-81 (Mart\'{\i} et al. 1993), IRAS~16547$-$4247 (Garay et al. 1996, Rodr\'{\i}guez et al. 2005), W3(H$_{2}$O) (Wilner et al. 1999), and L778-VLA~6 (Girart et al. 2002). It is usually proposed that these negative spectral indices are related to the presence of synchrotron emission from these objects. The synchrotron nature in a YSO jet was confirmed in the case of HH 80-81, where polarized emission was detected through sensitive radio observations at 6 cm (Carrasco-Gonz\'alez et al. 2010).
 
 Synchrotron emission implies the presence of relativistic particles. However, jets from YSOs are launched at relatively low velocities (in comparison with the jet velocities in AGNs), of the order of only several hundreds of kilometers per second. Therefore, an acceleration mechanism should be taking place in these objects in order to attain a population of relativistic particles that could produce detectable synchrotron emission. One possibility is that the acceleration of particles takes place in strong shocks where the jet impacts against the ambient medium. In this situation, particles could gain energy by diffusing back and forth across a shock front (e.g., Drury 1991). This process, known as diffusive shock acceleration (DSA), allows that particles originally moving at a few hundreds of kilometers per second can reach relativistic velocities. The DSA mechanism is known to work in AGN jets (e.g., Blandford et al. 1982), supernova remnants (e.g., Castro \& Slane 2010), nova ejecta (e.g., Kantharia et al. 2014), and colliding wind binaries (e.g., de Becker 2007), where shocks with velocities of at least several thousand kilometers per second are present. In this sense, protostellar jets could be a new extreme testing ground for DSA theory.

 The Serpens star-forming region is located at a distance of $\sim$415~pc (Dzib et al. 2010) and contains one of the first radio jets proposed to be a synchrotron emitter. This source, called the Triple Radio Source in Serpens, has a morphology consisting of a central thermal radio continuum source (positive spectral index), and two outer non-thermal lobes (negative spectral indices). Moreover, the outer knots show proper motions of the order of a few hundred kilometers per second, suggesting they are tracing out the motion of the jet against the ambient medium (Rodr\'{\i}guez et al. 1989, Curiel et al. 1993). In this paper, we present an analysis of new and archive data at radio wavelengths of the Triple Radio Source in Serpens, and discuss the results in the context of particle acceleration and synchrotron emission production. 
 
 \section{Observations}
 
 Observations of the Triple Radio Source in Serpens were made with the Karl G. Jansky Very Large  Array (VLA) of the National Radio Astronomy Observatory (NRAO)\footnote{The NRAO is a facility of the National Science Foundation operated under cooperative agreement by Associated Universities, Inc.}. We observed the continuum emission in the S, C, and X bands in B configuration during June 12 and 16, 2012 (project code: 12A-240). For each band, we observed a total continuum bandwidth of 2 GHz covering the frequency ranges 2-4 GHz, 4.5-6.5 GHz, and 8-10 GHz, in S, C, and X bands, respectively. Each band is divided in 1024 channels of 2 MHz. Bandpass and flux calibration were made by observing 3C286. Complex gain calibration was achieved by observation of 1824+1044 every 10 minutes. We also performed polarization calibration by using 3C286 as polarization angle calibrator and the unpolarized source 2355+4950 as leakage calibrator. The phase center of our observations was $\alpha$(J2000)=18$^h$29$^m$49.8$^s$, $\delta$(J2000)=$+$01$^{\circ}$15$'$20.6$''$. 
 
 We additionally analyzed VLA archive data taken at C band in the A configuration at 8 epochs spanning 18 years, from 1993 to 2011. In these observations, flux calibration was achieved by observing 3C286, while phase calibration was performed by using 1751+096. 
 
 Calibration of the data was undertaken with the data reduction package CASA (Common Astronomy Software Applications\footnote{https://science.nrao.edu/facilities/vla/data-processing}; version 4.1.0) following standard VLA procedures. Cleaned images were made using the task \emph{clean} of CASA. For the 2 GHz bandwidth images we used multifrequency synthesis (parameter \emph{nterms}=2) and multiscale cleaning (Rau \& Cornwell 2011). For these data, we made images selecting different bandwidths: 512 MHz, 2 GHz, and a single image using all three bands (S, C, and X). We used different values of the parameter \emph{robust} of \emph{clean}, ranging from $-$2 (uniform weighting) to $+$2 (natural weighting). For the analysis of the multiepoch archive data, we made images for each epoch by using parameter \emph{nterms}=1 and also different values of the \emph{robust} parameter. In order to better compare the images from the different epochs we convolved all the images to the same beam size. 
 
 A summary of the observations and image parameters are shown in Table \ref{tbl-obs} and \ref{tbl-images}, respectively.

\section{Results}
 
 In Figure \ref{SI_image} we show the continuum image (contours) obtained by using all the B configuration data in S, C and X bands, as well as the spectral index map (color scale) obtained from the multifrequency synthesis cleaning. In this figure, the jet-like morphology of the Serpens triple source is clearly seen. We identify four compact components: a central elongated source (C) and three outer knots (NW, NW\_C and SE). The three components C, NW\_C, y NW are connected by extended emission, whereas no similar extended emission is detected connecting the central source to the SE component. The SE knot appears splitted in two different components labeled as SE\_N and SE\_S, clearly seen in the higher angular resolution images (see below). In the following we discuss the characteristics of the radio jet.                                                                                                                                                                                                                                                                                                                                                                                                                      
 
\subsection{Spectral Indices and Spectral Energy Distributions} \label{SI}

\begin{figure}
 \epsscale{1.0}
 \plotone{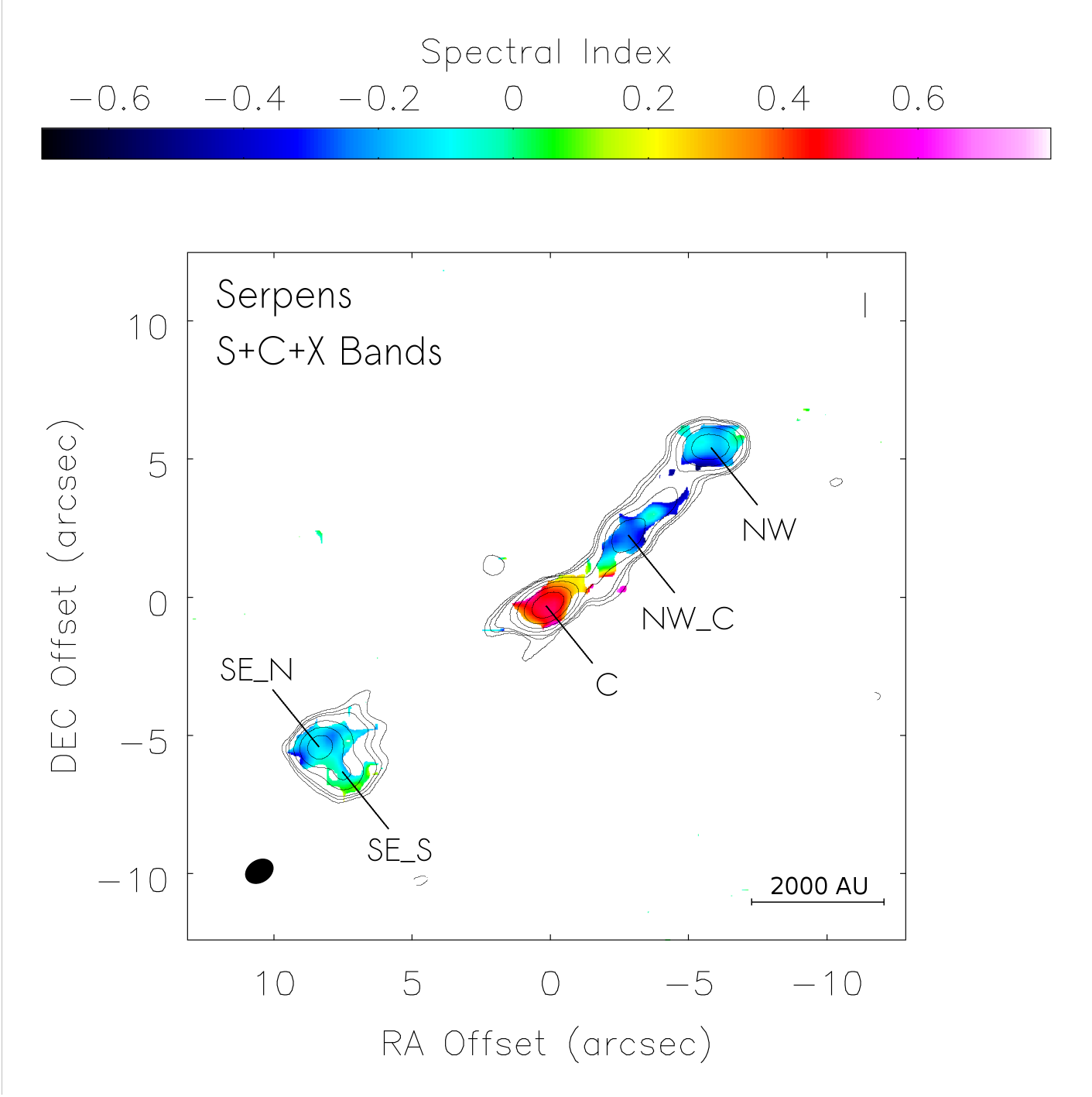}
 \caption{\footnotesize{Superposition of a radio continuum image made by combining data from S, C, and X bands of epoch 2012 
 (contours) over the spectral index image obtained from multifrequency synthesis cleaning (color scale). 
 Contours are $-$4, 4, 6, 8, 16, 31, 64, and 128 times the rms of the continuum image, $6\mu$Jy/beam. 
 The synthesized beam size is 1.07$''$$\times$0.80$''$ with a PA of $-$58$^\circ$. 
 The pixels shown in the spectral index image are those with an error in the spectral index of less than 0.1. 
 We labeled the four components discussed in the paper (central source C, and the outer knots, SE\_N, SE\_S, NW and NW\_C).}}\label{SI_image} 
 \end{figure}

\begin{figure}
 \epsscale{1.0}
 \plotone{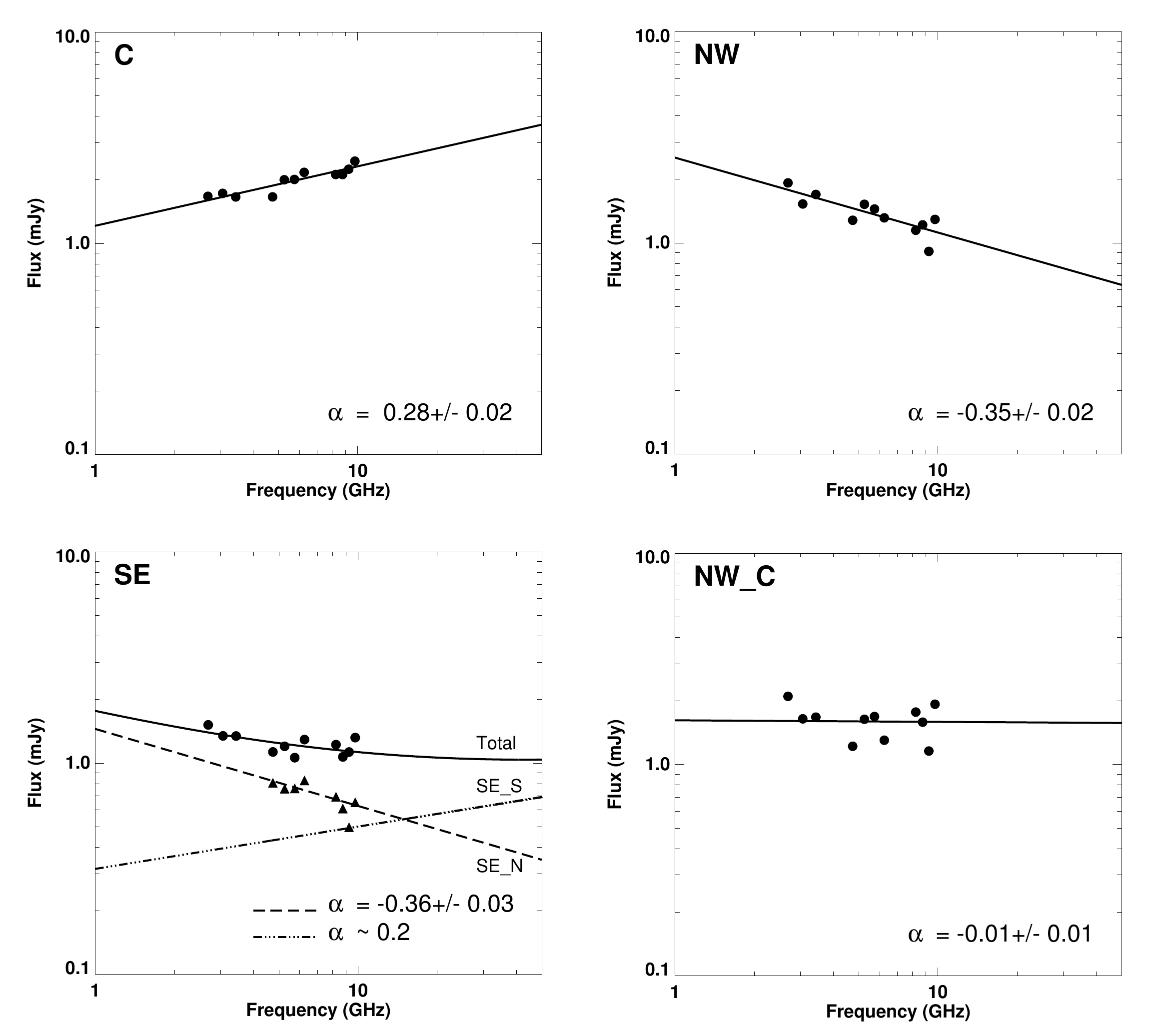}
 \caption{\footnotesize{Spectral energy distributions for the radio knots in the Serpens triple source as observed in 
 epoch 2012 (Table \ref{tbl-images}). 
 Flux densities are obtained from Gaussian fits to 512 MHz bandwidth natural weighted images convolved to the same beam size (2$\farcs$60).
 Solid lines in the C and NW panels are linear least-squares fits to the log data.
 In the SE panel, circles are total flux densities of the SE knot (SE\_N~+~SE\_S) obtained from low angular resolution images convolved
 to the beam size of the S band (2$\farcs$60). 
 Triangles are flux densities of the SE\_N component alone obtained from the higher angular resolution C and X band 
 images (uniform weighting; beam sizes=1$\farcs$40); these high angular resolution data suggest a negative spectral index for the 
 SE\_N source.
 The dashed line is a least-squares fit to these high angular resolution data, which show a negative spectral index.
 The dot-dashed line with a positive spectral index, corresponds to the SED of the SE\_S component, obtained as the difference 
 between the total flux of the SE knot and the flux of the SE\_N component. The NW\_C panel shows a linear least-squares fit to the 
 log data for the NW\_C knot, obtaining a flat spectrum.}} \label{sed}

 \end{figure}

 In order to study the nature of the radio emission in the triple source in Serpens, we obtained a spectral index map (Figure \ref{SI_image}), and the spectral energy distribution (SED) for each of the compact components in the jet (Figure \ref{sed}). Both the spectral index map and the SEDs show a difference between the nature of the emission of the central component, the lobes, and the extended emission. The central source shows a clear positive spectral index ($\sim$0.3), suggesting partially optically thick free-free emission and in good agreement with those of thermal radio jets (e.g. Anglada et al. 2015, Carrasco-Gonz\'alez et al. 2015). In contrast, the rest of the emission shows flat ($\alpha\sim$0) or negative spectral indices, which suggests optically thin free-free emission and non-thermal emission, respectively.
 
 The spectral index of NW is $-$0.35$\pm$0.02 and implies a non-thermal origin of the emission. The SED of the SE knot is more difficult to interpret. We know from the higher angular resolution images that this knot is composed by two radio knots (SE\_N and SE\_S, separated by $\sim$1.2$''$). However, in the lower frequency images, the angular resolution ($\sim$2$\farcs$6) is not high enough to separate the emission of these two components. The SED of this source obtained with low angular resolution appears very flat (see Figure \ref{sed}, left bottom panel). However, if we consider the flux densities obtained in the higher angular resolution images (C and X bands, uniform weighting; beam size = 1.4$''$), we obtain a negative spectral index ($-$0.36$\pm$0.03) for the SE\_N component. We then assume that the SED obtained from the low angular resolution images of SE can be decomposed in a non-thermal component (corresponding to SE\_N) plus a thermal component (most probably arising from SE\_S) (Figure \ref{sed}). For component NW\_C we also obtain a flat spectrum (Figure \ref{sed}). However, from the spectral index image of Figure \ref{SI_image}, which has higher angular resolution than the images used to construct the SED, we see that, at the position of the peak of NW\_C we obtain a negative spectral index ($\sim-$0.3). This image suggests that the contribution from optically thin free-free emission arises from the extended emission that surrounds component NW\_C, and could not be separated in the low angular resolution images, from which the SED shown in Figure \ref{sed} has been determined. We therefore conclude that the NW\_C component has also a non-thermal nature.
    
\subsection{Polarization} 

 Motivated by the detection of linearly polarized emission from the HH80-81 jet (Carrasco-Gonz\'alez et al. 2010), we carried out a  polarization study with our new S, C, and X VLA data. These are the most sensitive data obtained so far in the Serpens region that allow polarization calibration. However, we did not detect linearly polarized 
 emission from regions with negative spectral indices in the Serpens triple radio source (in none of the 2 GHz images we made). In Table~\ref{tbl-gp}, we give upper limits for the polarization degree in the different non-thermal radio knots implying that, if linearly polarized emission is present, its polarization degree should be less than 10\%. 

 There are two possibilities for this low polarization degree. First, an important difference between HH 80-81 (with polarization degrees $\sim 10-30$\%) and the triple source in Serpens is that synchrotron emission from the former seems to be detected in a large portion of the radio jet, from 0.1 to 0.5~pc from the protostar (see Carrasco-Gonz\'alez et al. 2010, 2013). It is then likely that most of the polarized emission is arising from a very collimated jet, where the magnetic field is expected to be well ordered. Indeed, the magnetic field lines in HH 80-81 seem to be parallel to the direction of the jet (Carrasco-Gonz\'alez et al. 2010). A well-ordered magnetic field would result in a relatively high polarization degree. In contrast, in Serpens, synchrotron emission seems to be detected mainly in the shocks against the ambient medium, where it is likely that the jet material is more turbulent. This could yield to a magnetic field very disordered in these shocks, and could result in a very low polarization degree. A second possibility is that the electron density in these shocks is high enough to result in a strong Faraday rotation of the polarization angle, resulting in a very low polarization degree when observed in a wide band. If this is the case, imaging with a smaller bandwidth
 (as it is was the case in the HH 80-81 of Carrasco-Gonz\'alez et al. 2010) would allow to detect higher polarization degrees. However, our observations of the Triple Radio Source in Serpens were very short in time, and the high sensitivity comes from the fact that we are using wide bandwidths. Imaging of smaller frequency ranges does not allow to obtain high sensitivity images to explore this possibility. Observations with a much larger observation time should be necessary in order to investigate if a strong Faraday rotation is present.
 
\subsection{Proper Motions}\label{pm}      
 
\begin{figure}
  \epsscale{0.4}
  \plotone{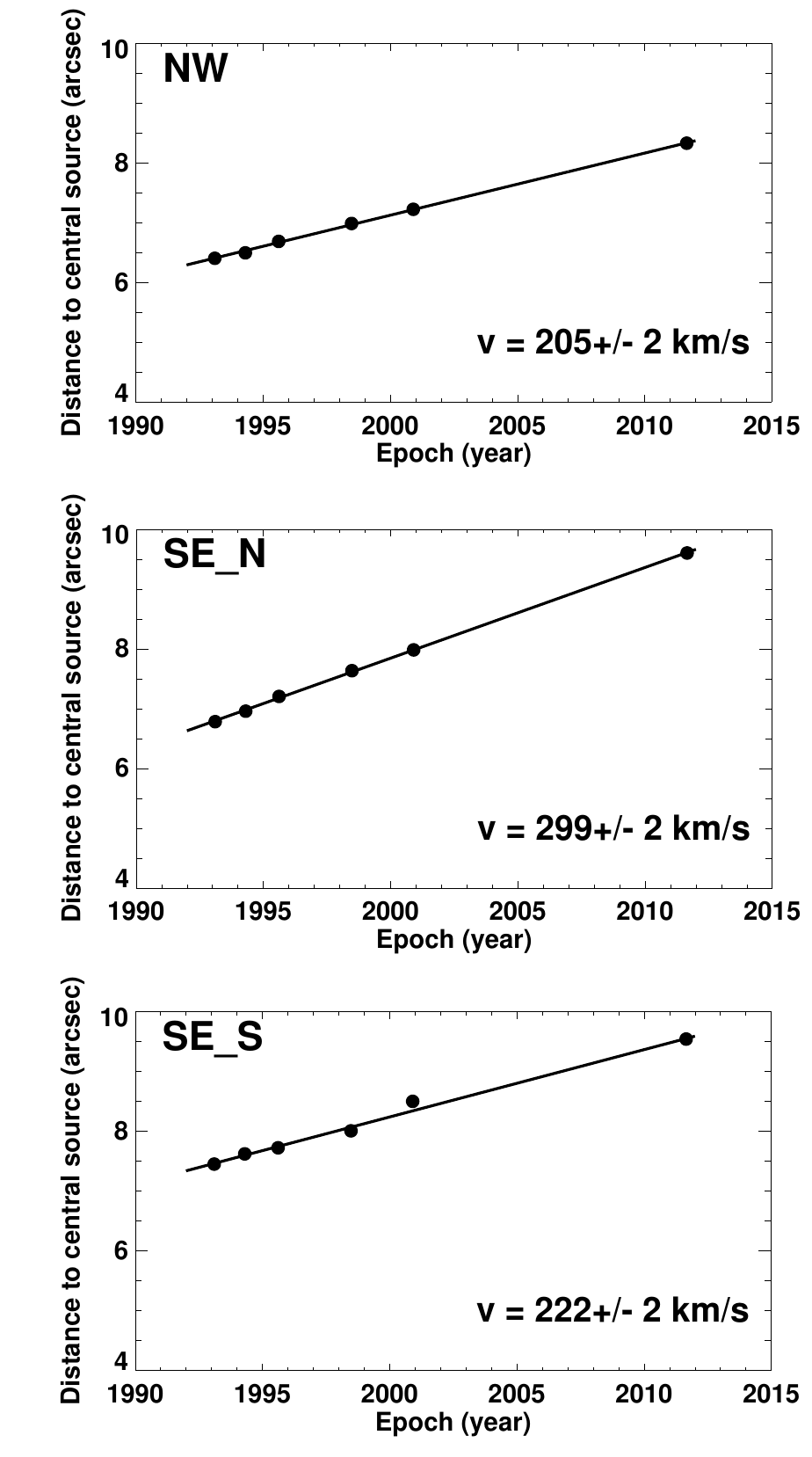}
  \caption{\footnotesize{Position vs time diagrams for the SE\_N, SE\_S and NW knots in the Serpens radio jet. 
  Positions are distances to the central source, which is assumed to remain at the same position at all epochs. 
  The solid lines are least squares fits to the data. Velocities obtained in each fit are labeled 
  in the panels. \label{pos_time}}}
  \end{figure}

\begin{figure}
  \epsscale{0.5}
  \plotone{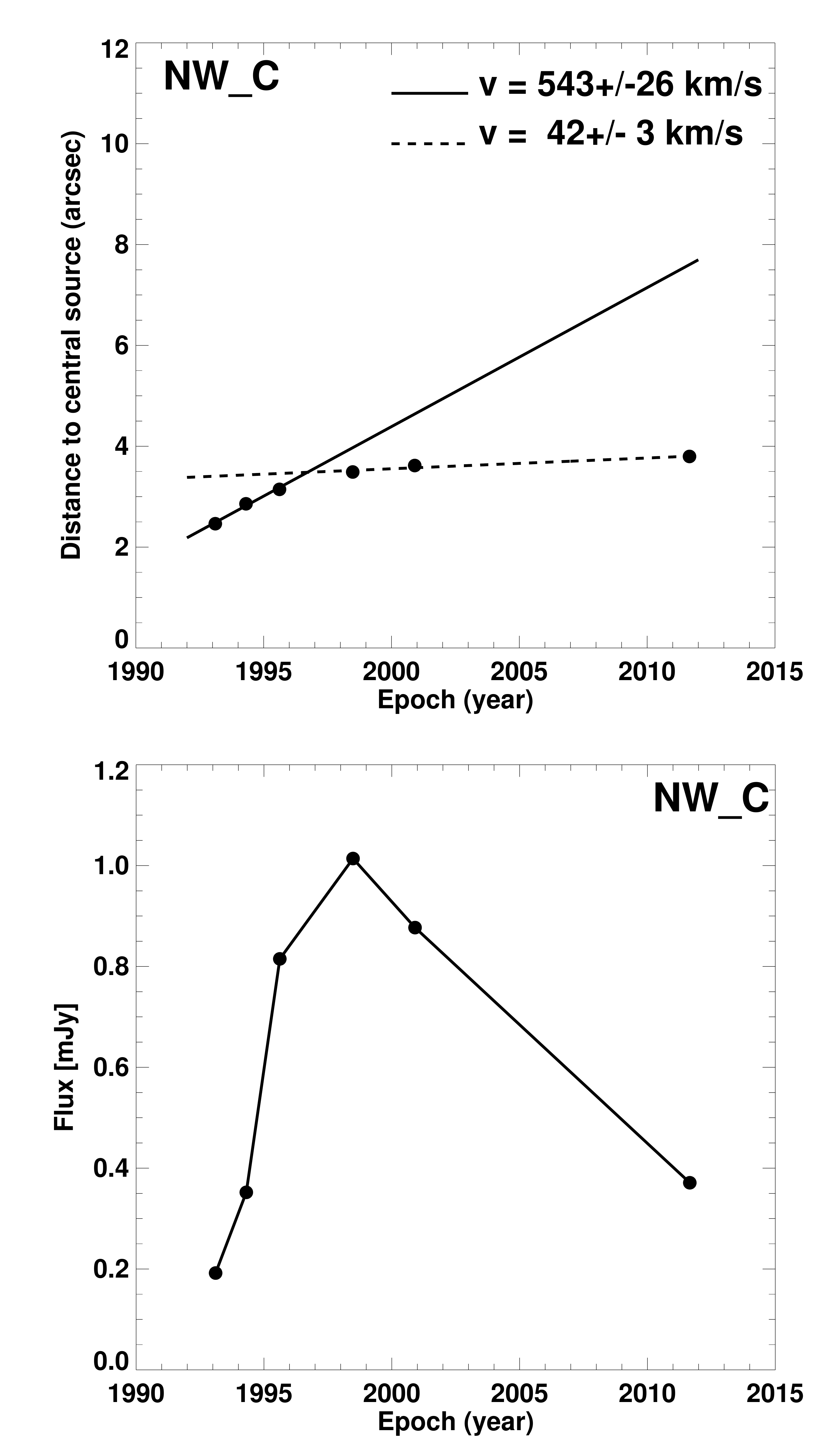}
  \caption{\footnotesize{Upper panel: Position vs time diagram for the NW\_C knot. Lines are least squares fits to the data before 
  1998 (solid lines) and after 1998 (dashed line). 
  This knot shows two different velocities: between 1993 and 1998, the knot seems to move with a constant velocity 
  of $\sim$500 km~s$^{-1}$, then, it decelerates and moves at a lower velocity of $\sim$40 km~s$^{-1}$. 
  Lower panel: A light curve of the flux density at C band. 
  The flux density rises until 1998, when it reaches its maximum and then decays.  \label{flux_time}}}
 \end{figure}

 We studied the kinematics of the different knots in the Serpens radio jet by analysing the multiepoch high angular resolution archive data at 6 GHz (see Table \ref{tbl-images} for multi-epoch image parameters). We aligned the images obtained at different epochs by assuming that the central source has the same position in each of the observed epochs. In Figure \ref{int_all}, we show in three columns, multiepoch images for the observed knots. The images are all shown with the same intensity scale to compare the fluxes at different epochs. We see that all knots are moving away from the central source. Measuring their displacements in each epoch and assuming a distance of 415 pc to the Serpens molecular cloud, we estimated the tangential (i.e., in the plane of the sky) velocities of the knots (Figure \ref{pos_time} and Table 4). We found that the NW and SE\_S tangential velocities are similar ($\sim$200 km~s$^{-1}$), while the SE\_N is moving faster ($\sim$300 km~s$^{-1}$). The velocities of these three knots seem to be constant during the analyzed epochs. On the other hand, we observe an interesting behaviour in the NW\_C knot: it moves away from the protostar with a very high velocity $\sim$500 km~s$^{-1}$ between 1993 and 1998, and after several years it dramatically decelerates to a velocity of only $\sim$40 km~s$^{-1}$. Furthermore, the flux density of NW\_C also varies with time, increasing when it moves fast, and decreasing, when the velocity is low (see Figure \ref{flux_time}). Recent observations at 7~mm performed by Choi (2009) detected a dusty filament-like structure that, in projection, seems to pass across the jet at the position where NW\_C decelerates (see Figure \ref{filamento}). Therefore, we speculate that the NW\_C decelerates after interacting with this dusty filament. In this case, it would be expected an increase in the plasma density and therefore in the flux density.
     
\begin{figure}
  \epsscale{1.0}
  \plotone{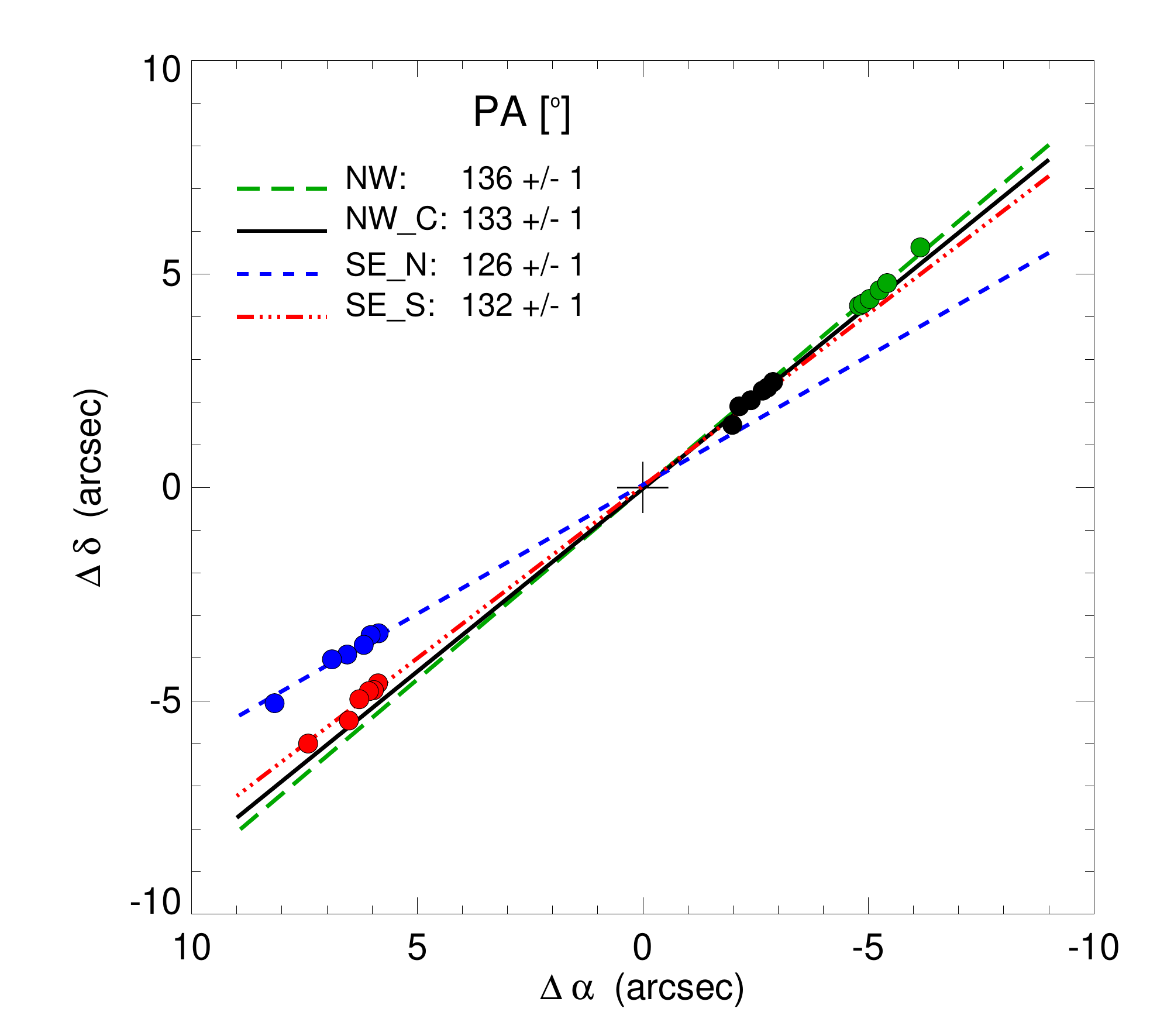}
  \caption{\footnotesize{A diagram showing the positions of all the knots in all analyzed epochs. 
  Lines are least squares fits to the positions of the knots in the different epochs.
  We assume that knots arised from the central source. 
  Therefore, we included the position of the central source in each of the fits. \label{PA}}}
 \end{figure}

 In Figure \ref{PA} we show the positions of the knots relative to the central source for the six epochs between 1993 and 2011, as well as least squares fits to the trajectories. The position angles (P.A.s) are derived for the motion of each component. Components NW, NW\_C and SE\_S seem to be moving with similar P.A.s in the range 132-136 degrees. In contrast, component SE\_N seems to move in a different direction with a P.A. of 126 degrees. We also note that the central source appears elongated in all epochs with a PA of 119 $\pm$ 1 degrees, which is closer to the value obtained for the motion of SE\_N.
 
 We also estimated the kinematic ages of the different knots, i.e. the time needed for they to move from the central source to their present position. We assume that SE\_N, SE\_S and NW moved with constant velocity since they emerged from the central source. For NW\_C, we adopt a constant velocity equal to the value measured before it decelerates in 1998. We found similar kinematic ages of $\sim$80 years for NW and SE\_S, which suggests they both arised at the same time in $\sim$1930. This is also consistent with these two knots showing similar velocities ($\sim$200 km~s$^{-1}$) and a similar direction of their movement (P.A.$\simeq$130$^\circ$). For SE\_N, we found a kinematic age of $\sim$60 years, suggesting it arised from the central source later, around $\sim$1950. The P.A. of this knot (126$^\circ$) is different from that of the SE\_S and NW knots, which suggests that the jet suffered a change in its P.A. of $\sim$10$^\circ$ in $\sim$20~years. The youngest knot is NW\_C, which we estimate was ejected from the central source around $\sim$1980. This knot moves in a direction with a P.A. similar to that of the earlier ejected knots SE\_S and NW. This suggests that the jet changed again its direction between 1950 and 1980, to an orientation similar to that of 1930. Finally, the central source is presently elongated approximately in the direction of the SE\_N knot. This behaviour is consistent with precession of the jet and an episodic ejection phenomenon every 20-30 yr. If periodicity of these strong ejecta were confirmed, it would suggest the presence of a close companion orbiting around the driving source of the radio jet. Precession of the jet axis could be driven by tidal interactions between the disk from which the jet originates and a companion star in a non-coplanar orbit (Masciadri \& Raga 2002; Anglada et al. 2007). 

 Several authors have proposed that the central source of the Triple Radio Source in Serpens is actually a binary system of YSOs (e.g., Eiroa \& Casali 1989; Hodapp 1999; Eiroa et al. 2005; Choi 2009; Dionatos et al. 2010). Indeed, Dionatos et al. (2014) found evidence of the existence of a binary companion, lying at ~1$\farcs$5 to the NW, which corresponds to a separation in the plane of the sky of $\sim$622~AU. We investigated if this companion could be responsible of the observed jet precession in our VLA data through tidal interactions. Following the equations presented in Anglada et al. (2007) it is possible to infer the separation of the binary system from the precession of the jet. Hence, assuming a period of 20-30~years, and $\beta=$10$^\circ$ (the angle between the central flow axis and the line of maximum deviation of the flow from this axis), the separation between the components of a binary system responsible for the precession should be $\sim$3~AU. Therefore, if the observed ejections have a binary system origin, the companion orbiting around the driving source of the radio jet should be much more close than the companion detected by Dionatos et al. (2014). This would require of very high angular resolution observations to be confirmed. Furthermore, if the source reported by Dionatos were responsible for the precession, it should be located roughly perpendicular to the jet axis.
 
\section{Discusion: On the particle acceleration  and synchrotron emission}\label{acc}
 
 Protostellar jets usually show a simple morphology at radio frequencies, consisting of an elongated source with a positive spectral index interpreted as free-free emission from ionized material tracing out the base of the large-scale jet (Anglada et al. 1998). The triple source in Serpens also shows this central free-free radio source. However, as in the case of other protostellar radio jets, such as HH 80-81, it also shows non-thermal radio knots located at larger distances from the central protostar. The most likely scenario to explain the presence of these non-thermal radio knots is that they are tracing strong shocks against the ambient medium where it is possible to accelerate particles that emit synchrotron radiation. However, this phenomenon only seems to be possible in some protostellar jets. In the following, we discuss about the physical properties of the triple radio source in Serpens and their relationship to the particle acceleration phenomenon. 
 
\begin{figure}
  \epsscale{0.8}
  \plotone{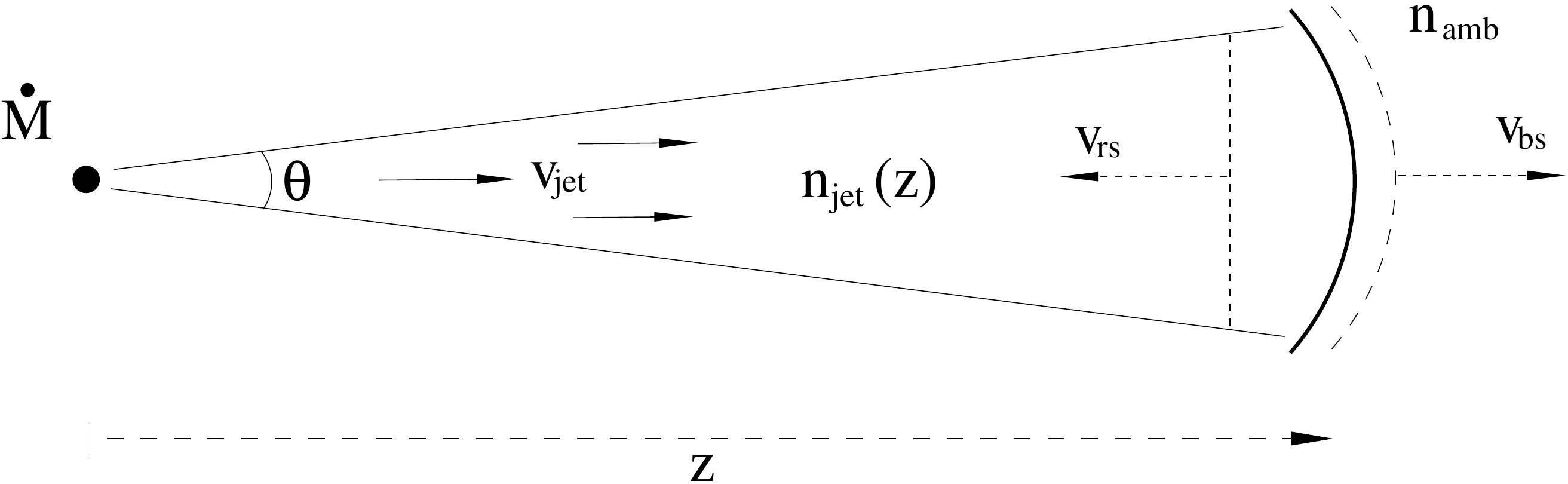}
  \caption{\footnotesize{Scheme showing the jet paramenters involved in Equations \ref{vrs} - \ref{densidad}. \label{esquema}}}
 \end{figure}

 The intriguing aspect of particle acceleration in protostellar jets is to understand how these relatively slow jets are able to accelerate particles up to relativistic velocities. One possibility is that these non-thermal protostellar jets are particularly fast. Indeed, in the case of HH 80-81, proper motions of internal shocks suggest jet velocities of $\sim$1000~km~s$^{-1}$ (Mart\'{\i} et al. 1995). These velocities are considerably large compared with typical velocities in protostellar jets, of the order of a few hundreds of km~s$^{-1}$ (e.g., Rodr\'{\i}guez et al. 2000, Estalella et al. 2012). In the case of Serpens, there are no observations with an angular resolution high enough to measure the velocity of the jet material as it emerges from the protostar. However, we have measured proper motions of $\sim$200-300~km~s$^{-1}$ of non-thermal ($\alpha \sim -0.35$) radio knots located at $z\simeq 9\arcsec \simeq 0.02~{\rm pc}$ from the central source. We interpret these knots as synchrotron emission produced where the jet impacts against the ambient medium (see Fig.~\ref{esquema}). Therefore, if the molecular cloud is denser than the jet at the position of the shock, the jet velocity should be larger than the velocities observed in the synchrotron radio knots since the material of the jet should slow down in the shock.

 Synchrotron emission at 6~cm is produced by relativistic electrons with Lorentz factors $\gamma_6 \sim 60(B/\rm mG)^{-1.5}$ in a magnetic field $B$. These  particles can be accelerated in the bow shock with the ambient medium or in the jet reverse shock (Mach disk). The acceleration mechanism depends on the nature of the shocks, radiative or adiabatic (i.e. non-radiative). A way to discern whether the shocks are radiative or adiabatic is by comparing the thermal cooling distance $d_{\rm cool}$ with the radius of the jet at the position of the shock, $r_{\rm jet}$ (Blondin et al. 1989). The cooling distance can be estimated as:

 \begin{eqnarray}
 d_{\rm cool} =  2\times10^{13} \left(\frac{n}{100~{\rm cm^{-3}}} \right)^{-1} \left(\frac{v_{\rm s}}{100~{\rm km~s^{-1}}} \right)^{-4.51}~{\rm cm} & ; & v_s < 80~km~s^{-1} \label{cool1}\\
 d_{\rm cool} =  1.7\times10^{14} \left(\frac{n}{100~{\rm cm^{-3}}} \right)^{-1} \left(\frac{v_{\rm s}}{100~{\rm km~s^{-1}}} \right)^{4.73}~{\rm cm} & ; & 80 < v_s < 400~km~s^{-1} \label{cool2}\\
 d_{\rm cool} =  2.24\times10^{14} \left(\frac{n}{100~{\rm cm^{-3}}} \right)^{-1} \left(\frac{v_{\rm s}}{100~{\rm km~s^{-1}}} \right)^{4.5}~{\rm cm} & ; & v_s > 400~km~s^{-1}  \label{cool3}
 \end{eqnarray}
 
 \noindent where \emph{n} is the density of the medium where the shock is propagating, and $v_s$ is the velocity of the shock. In the case of Serpens, we can estimate $r_{\rm jet}$ by taking half of the beam size of the highest angular resolution observations presented here, since the shocks appear unresolved, i.e., $r_{\rm jet} \simeq 0\farcs235 \simeq 1.5\times10^{15}~{\rm cm}$. In what follows we discuss the conditions required to accelerate electrons up to $\gamma_6$ in the bow shock and in the Mach disc.

\subsection{Synchrotron emission from the shocked molecular cloud}

 Assuming that the jet is in the plane of the sky, the bow shock velocities are $v_{\rm bs} \simeq 200-300$~km~s$^{-1}$, as obtained from our proper motions  analysis. Therefore, we calculate the cooling distance with equation \ref{cool2} assuming that the density \emph{n} is the ambient density ($n=n_{amb}$), and the shock velocity is the measured bow-shock velocity ($v_{s}=v_{bs}$). The density of the molecular cloud, obtained through ammonia emission, is estimated as $n_{\rm amb} \sim 4\times 10^{4}$~cm$^{-3}$ (Curiel et al. 1996). Then, we find $d_{\rm cool,bs}/r_{\rm jet} = 0.02$ which implies that the shocks against the molecular cloud are radiative. In this situation, the flat spectral index detected in the radio knots can be explained as acceleration and compression of cosmic rays in the molecular cloud, as was suggested for old supernova remnants (i.e. in the radiative phase) emitting synchrotron radiation (Chevalier 1999). Another possibility is that electrons are accelerated through second order Fermi acceleration (Ostrowski 1999).

\subsection{Synchrotron emission from the shocked jet}

 In order to study the conditions in the reverse shock, we need to know the jet parameters. The velocity of the reverse shock can be estimated as
 
\begin{equation}\label{vrs}
v_{\rm rs}=v_{\rm jet}- 3v_{\rm bs}/4, 
\end{equation}

\noindent while the jet velocity is given by (Raga et al. 1998)

\begin{equation}\label{vjet}
\frac{v_{\rm jet}}{v_{\rm bs}}=\frac{(1+\beta)}{\beta}, \hspace{20pt} 
\beta=\sqrt{\frac{n_{\rm jet}}{n_{\rm amb}}}. 
\end{equation}

 For a pure hydrogen conical jet with opening angle $\theta$ and a mass-loss rate $\dot{M}$, the jet density  is given by (Reynolds 1986)
 
\begin{equation}\label{densidad}
\left(\frac{n_{\rm jet}}{\rm cm^{-3}} \right)= 
\frac{3.95 \times 10^{7}}{4\pi(1-\cos\theta/2)} 
\left(\frac{\dot{M}}{\rm{M_{\sun}~yr^{-1}}} \right) 
\left( \frac{v_{\rm jet}}{\rm{km~s^{-1}}} \right)^{-1}
\left( \frac{z}{\rm pc} \right)^{-2}.
\end{equation}

 For the Serpens radio-jet we estimate an opening angle of $ \theta\sim r_{\rm jet}/z \sim 3^{\circ}$, being z the distance from the central source to the shocks. The jet velocity, jet density and mass-loss rate are unknown parameters. However, since all the jet parameters are related through the previous equations, we explored different possibilities. We assumed different combinations of $10^{-7} \le \dot{M} \le 10^{-5}$~M$_{\sun}$~yr$^{-1}$ and $400 \le v_{\rm jet} \le 1200$~km~s$^{-1}$. Then, given $v_{jet}$ and $\dot{M}$, we calculate $n_{\rm jet}$, $v_{\rm bs}$ and $v_{\rm rs}$ using equations \ref{vrs}-\ref{densidad}. The cooling distance is calculated with equations \ref{cool1}-\ref{cool3}, where the density \emph{n} is the jet density ($n= n_{jet}$), and the shock velocity equals the reverse-shock velocity ($v_{s}=v_{rs}$). In this way, we can study if a given pair of $v_{jet}$ and $\dot{M}$ results in an adiabatic or radiative reverse shock. 
 
\begin{centering}
  \begin{figure}
  \includegraphics[scale=0.5, angle=0]{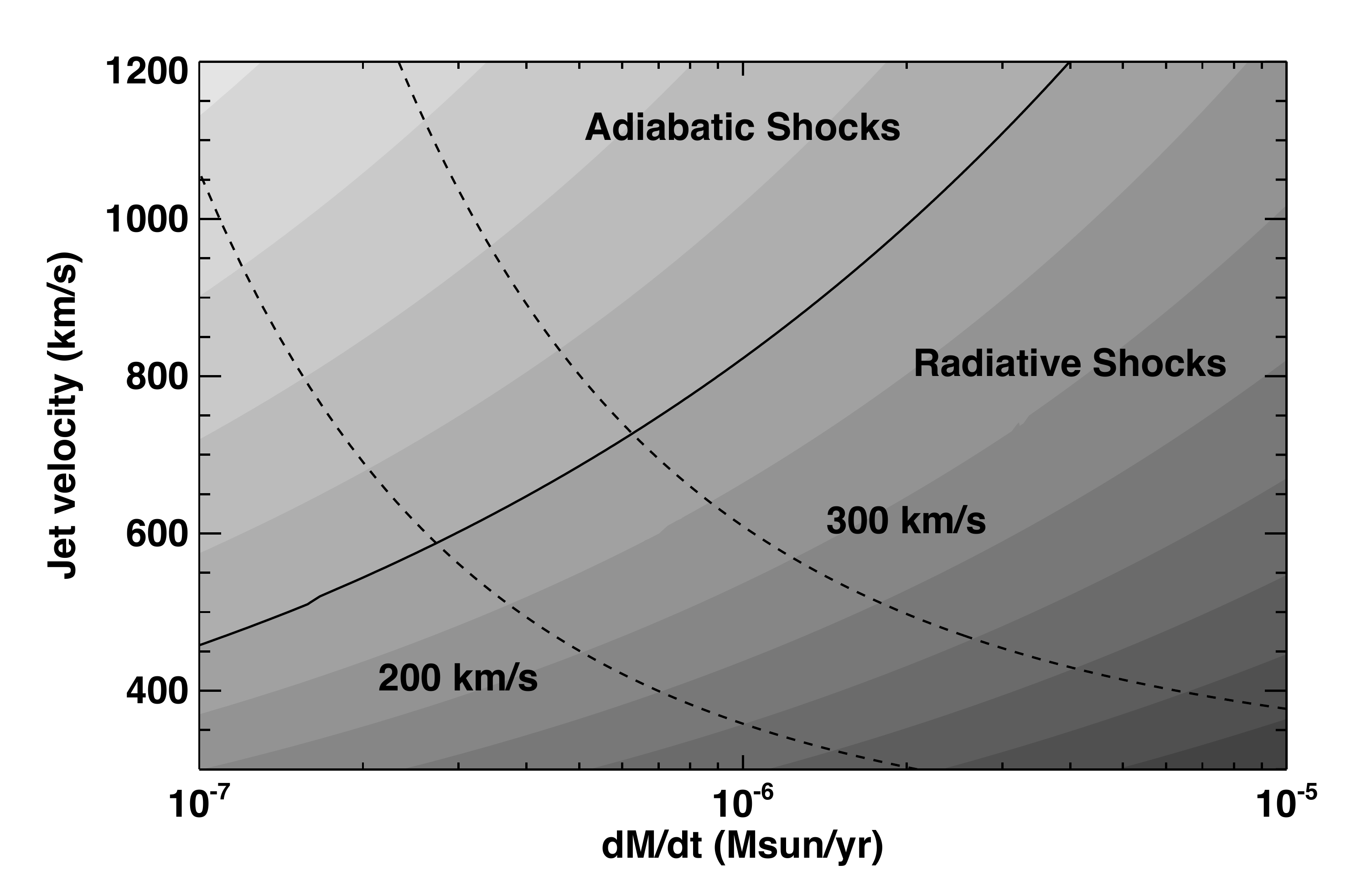}
   \caption{\footnotesize{We plot the ratio
  $d_{cool,rs}/r_{jet}$ (grey sacale) for different combinations of $\dot{M}$ and $v_{jet}$. The solid line separates
  the jet conditions that result in adiabatic shock (i.e. $d_{cool,rs} > r_{jet}$ ) from those that result in 
  radiative shocks. We also show two dashed lines corresponding to $v_{bs}$= 200 and 300 km s$^{-1}$. We can see that in order to be the Mach disk an adiabatic 
  shock, a mass-loss rate $\dot{M}\lesssim 5 \times 10^{-7} \rm M_{\sun}$~yr$^{-1}$ and ${v_{\rm jet}\gtrsim500~{\rm km~s^{-1}}}$ are needed. \label{adiabatic}}}
  \end{figure}
 \end{centering}

 The results of the above procedure are shown in Figure~\ref{adiabatic}. In this Figure, we show a line that separates the combinations of $\dot{M}$ and $v_{\rm jet}$ that result in an adiabatic reverse shock (i.e. $d_{\rm cool,rs}> r_{\rm jet}$) from those that result in radiative reverse shocks. We also show two lines corresponding to $v_{\rm bs}$= 200 and 300~km~s$^{-1}$, the observed velocities of the bow-shocks. We can see that, in order for the Mach disk to be a non-radiative shock, and the bow-shocks to move with the observed velocities, the jet should have a mass-loss rate $\dot{M}\lesssim 5\times10^{-7}~{\rm M_{\sun}~yr^{-1}}$ while the jet material should move at velocities ${v_{\rm jet}\gtrsim500~{\rm km~s^{-1}}}$. This suggests that a jet with a typical mass-loss rate for an intermediate-mass protostar is able to accelerate particles via the DSA mechanism if the velocity of the jet is high enough.

 In such a case, electrons can be accelerated via DSA and injected in the shock downstream region following a power-law energy distribution $\propto \gamma^{-p}$, with $p = 2$.  This population of non-thermal electrons produces 
synchrotron emission with spectral index $\alpha = (1-p)/2 = -0.5$. 
 We note however that  $p \ne 2$  is also 
possible in oblique shocks given that the diffusion approximation breaks down 
depending on the inclination angle between the magnetic field and the shock 
normal (Bell et al. 2011). This effect can account for the spectral index
$\alpha\simeq-$0.35 measured in the Serpens radio knots, flatter than the 
typical value -0.5.  On the other hand, a flatter spectral index  may result 
from thermal ($\alpha>$0) emission contamination, increasing the flux at 6~cm 
and flattening the spectrum.

 From our analysis of the proper motions of the non-thermal radio knots we found that the jet is precessing and that non-thermal bow-shocks are excited only at certain epochs. This leads us to think that most of the time, the jet parameters do not meet the necessary conditions to produce efficiently relativistic particles in shocks. Since the bow-shocks seem to be excited periodically, we speculate that periodic interactions of the driving source of the jet with a close companion could increase the jet velocity (and maybe also, to produce a slight increase in the mass-loss rate).

\section{Conclusions}
  
We have presented an analysis of new and archive VLA observations of the triple radio source in Serpens. Our main conclusions can be summarized 
as follows:
  
\begin{itemize}

  \item The triple source in Serpens presents a clear jet-like morphology with a central source and several outer radio knots. The central source shows a positive spectral index at cm wavelenths consistent with partially optically thick thermal free-free emission, in well agreement with typical radio jets found in other YSOs. In contrast, the outer knots in the jet, tracing out the movement of the jet through the surrounding medium, show negative spectral indices. This suggests the presence of synchrotron emission, and consequently that a mechanism responsible for particle acceleration up to relativistic velocities might be taking place in these shocks. 
  
  \item We measured proper motions for the outer knots, and found projected velocities of 200-300 km~s$^{-1}$. These radio knots are most probably tracing shocks of the jet against the dense ambient medium. All this implies that the jet velocity should be high compared with typical velocities in other protostellar jets. 

  \item Linearly polarized emission is not detected in the radio knots showing negative spectral indices. We estimated upper limits for the polarization degree of $\sim$10\%. There are two possible explanations for this low polarization degree: (1) the magnetic field at the jet termination shocks is disordered, or (2) a high electron density in the shocks results in a strong Faraday depolarization of the synchrotron emission. 

  \item The change in direction of the proper motions of the radio knots suggest precession of the jet. Moreover, their kinematic ages also suggest they were ejected episodically every 20-30 yrs. These results could suggest the presence of a companion orbiting the driving source of the radio jet. If this is the case, we estimated a separation of $\sim$3~AU for this binary system. 

  \item Given that the synchrotron emitter remains unresolved in our 6 cm observations, we conclude that the emission arises in a compact region. Thus, the emission at 6~cm can be produced in the downstream region of the radiative bow shock, or in the jet shocked region. We discuss these two possibilities and conclude that particle acceleration via the Fermi I mechanism can take place only in the Mach disk under certain conditions. We found that the jet must eventually satisfy $\dot{M}\lesssim 5\times10^{-7}~{\rm M_{\sun}~yr^{-1}}$ and ${v_{jet}\gtrsim500~{\rm km~s^{-1}}}$ in order for the Mach disk to be a non-radiative shock and an efficient particle accelerator.
 
  \item Our results on the Triple Radio Source in Serpens indicate that particle acceleration via DSA could be possible in jets from intermediate-mass protostars. The jet does not need to show very extreme characteristics, as a high mas-loss rate, but it seems to be only necessary that the jet reaches moderately high velocities  ($\sim$600~km~s$^{-1}$) at certain epochs. Interactions with the proposed close binary companion could result in episodic increase of the jet velocity and the production of synchrotron emission in the shocks against the ambient medium. 
  
\end{itemize}

\begin{figure}
  \epsscale{0.7}
  \plotone{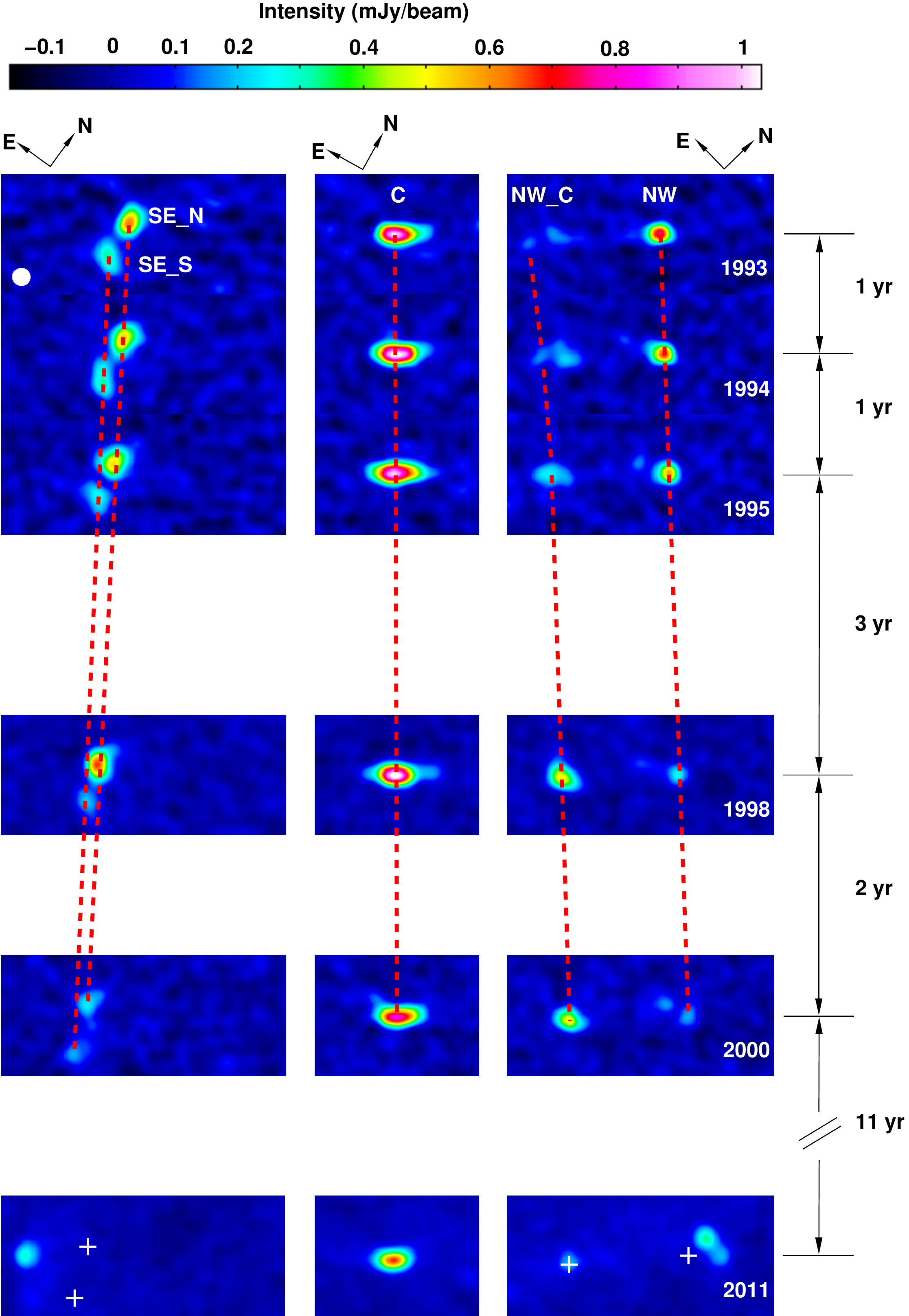}
  \caption{\footnotesize{VLA C-band, A configuration intensity images of the components of the Serpens radio jet in different epochs. 
  All images are convolved to the same circular beam size, 0$\farcs$47. 
  The comparison of the different images reveal proper motions of the outer knots (SE\_N, SE\_S, NW\_C and NW). 
  The separations between images of epochs from 1993 to 2000 are proportional to the separations in time. 
  Dashed red lines join the positions of the knots in different epochs up to 2000. White marks in the 2011 image correspond to the 
  positions of each knot at epoch 2000.
  \label{int_all}}}
 \end{figure}

\begin{figure}
  \epsscale{1.0}
  \plotone{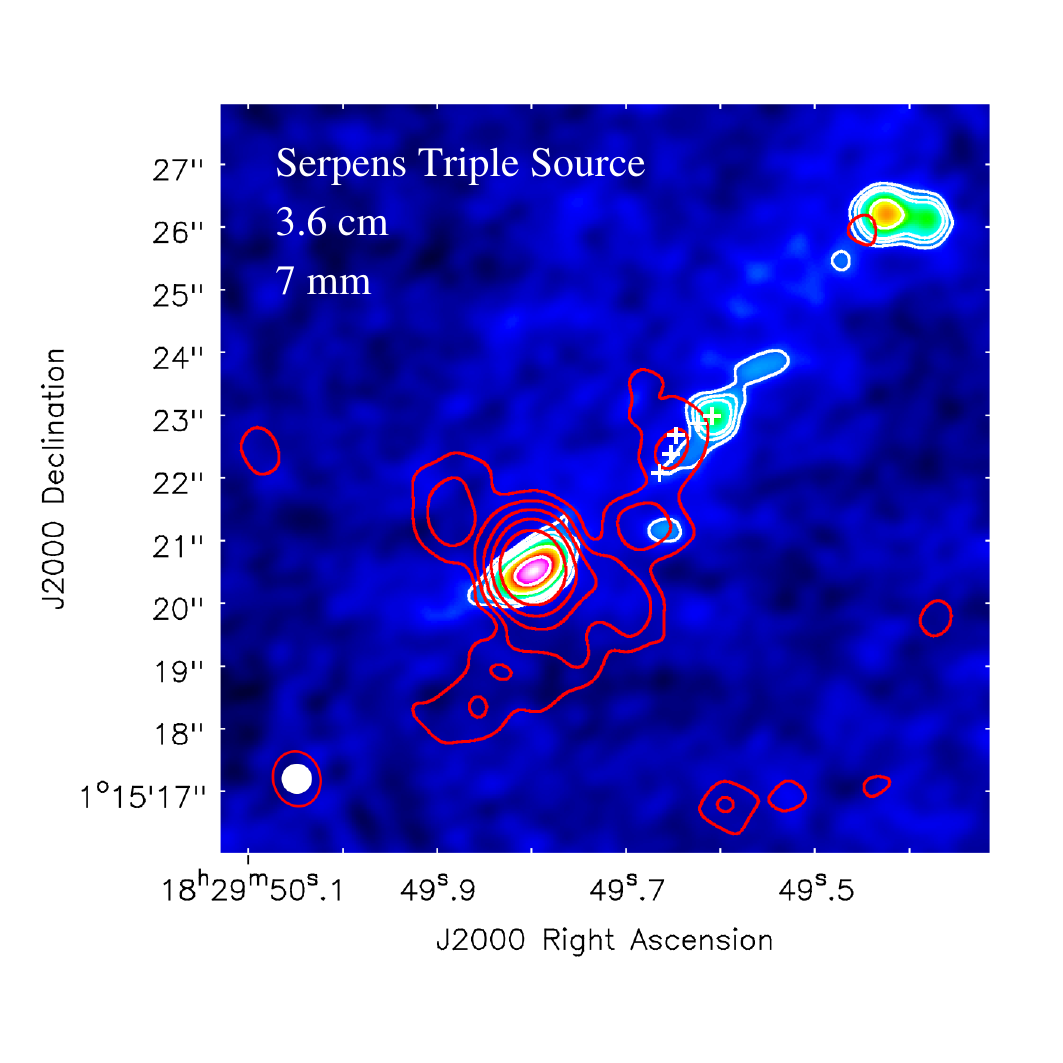}
  \caption{\footnotesize{Superposition of a radio continuum image at 7~mm (red contours) obtained from Choi (2009), over a radio continuum 
  image at 3.6~cm, corresponding to 2011 (color scale and white contours). White contours are 4, 6, 8, and 16 times the rms 
  noise (0.02 mJy~beam$^{-1}$), and the synthesized 
  beam size is 0$\farcs$47. Red contours are 3, 6, 12, 18, 24, and 48 times the rms noise, 0.05 mJy~beam$^{-1}$. Synthesized beam size is 0$\farcs$88 $\times$ 0$\farcs$75 with a PA of 15$^\circ$. The white marks 
  are the positions of NW\_C at the six epochs between 1993 and 2011.\label{filamento}}}
 \end{figure}

 A.R-K. and C.C-G. acknowledge support by UNAM-DGAPA-PAPIIT grant number IA101214 and financial support from MINCyT-CONACyT ME/13/47 grant, corresponding to the Bilateral Cooperation Program. G.A. and J.M.T. acknowledge support from MICINN (Spain) AYA2011-3O228-CO3 and AYA2014-57369-C3 grant (co-funded with FEDER funds). A.T.A. acknowledges support from the UK Science and Technology Facilities Council under grant No. ST/K00106X/1. 
 We also thank to Minho Choi and Gisela Ort\'iz-Le\'on for providing us data to perform figures \ref{filamento} and \ref{int_all}, respectively.

 \begin{deluxetable}{lccccc}
  \tabletypesize{\normalsize}
  \tablecaption{OBSERVATIONS \label{tbl-obs}}
  \tablewidth{0pt}
  \tablehead{
  \colhead{\centering Observation} & 
  \colhead{VLA} &  
  \colhead{Frequency} &  
  \colhead{Flux} & 
  \colhead{Phase} &  
  \colhead{Leakage} \\
  \colhead{\centering Date} &
  \colhead{Configuration} &
  \colhead{(GHz)} &
  \colhead{Calibrator} &
  \colhead{Calibrator} &
  \colhead{Calibrator} 
  }
  \startdata
  1993 Jan 09        &    A    &    6.0    &    3C 286    &    J1751+096    &  ---  \\  
  1994 Mar 21        &    A    &    6.0    &    3C 286    &    J1751+096    &  ---  \\  
  1995 Jul 13        &    A    &    6.0    &    3C 286    &    J1751+096    &  ---  \\  
  1998 May 09        &    A    &    6.0    &    3C 286    &    J1751+096    &  ---  \\  
  1998 May 30        &    A    &    6.0    &    3C 286    &    J1751+096    &  ---  \\  
  1998 Jun 01        &    A    &    6.0    &    3C 286    &    J1751+096    &  ---  \\  
  2000 Oct 27        &    A    &    6.0    &    3C 286    &    J1751+096    &  ---  \\  
  2011 Jun 26$^{a}$  &    A    &    6.0    &    3C 286    &    J1804+010    &  ---  \\  
  2012 Jun 12        &    B    &    3.0    &    3C 286    &    J1824+1044   &  J2355+4950  \\   
  2012 Jun 12        &    B    &    6.0    &    3C 286    &    J1824+1044   &  J2355+4950   \\  
  2012 Jun 12        &    B    &   10.0    &    3C 286    &    J1824+1044   &  J2355+4950   \\  
  2012 Jun 16        &    B    &    3.0    &    3C 286    &    J1824+1044   &  J2355+4950   \\  
  2012 Jun 16        &    B    &    6.0    &    3C 286    &    J1824+1044   &  J2355+4950   \\  
  2012 Jun 16        &    B    &   10.0    &    3C 286    &    J1824+1044   &  J2355+4950   \\  
  \enddata
   \tablenotetext{a}{See Ort\'{\i}z-Le\'on et al. 2015 for a full description on these observations.} 

 \end{deluxetable}

  \begin{deluxetable}{lccccc}
  \tabletypesize{\normalsize}
  \tablecaption{PARAMETERS OF THE IMAGES \label{tbl-images}}
  \tablewidth{0pt}
  \tablehead{
  \colhead{\centering Observation} & 
  \colhead{Spectral} &  
  \colhead{Configuration} &
  \colhead{Bandwidth} &  
  \colhead{Weighting} & 
  \colhead{Synthesized} \\
  \colhead{\centering Date} &
  \colhead{Band} &
  \colhead{ } &
  \colhead{(GHz)} &
  \colhead{ } &
  \colhead{Beam}
  }
  \startdata
  1993 Jan 09             &    C          &    A    &    0.1      &    Robust = 0      &    $0.47''$                   \\     
  1994 Mar 21             &    C          &    A    &    0.1      &    Robust = 0      &    $0.47''$                   \\
  1995 Jul 13             &    C          &    A    &    0.1      &    Robust = 0      &    $0.47''$                   \\
  1998$^{a}$              &    C          &    A    &    0.1      &    Robust = 0      &    $0.47''$                   \\
  2000 Oct 27             &    C          &    A    &    0.1      &    Robust = 0      &    $0.47''$                   \\ 
  2011 Jun 26             &    C          &    A    &    0.1      &    Natural         &    $0.47''$                   \\
  2012$^{b}$              &    S          &    B    &    2.0      &    Uniform         &    $2.60''$                   \\   
  2012$^{b}$              &    C          &    B    &    2.0      &    Uniform         &    $2.60''$                   \\      
  2012$^{b}$              &    X          &    B    &    2.0      &    Uniform         &    $2.60''$                   \\
  2012$^{b}$              &    C          &    B    &    2.0      &    Natural         &    $1.40''$                   \\      
  2012$^{b}$              &    X          &    B    &    2.0      &    Natural         &    $1.40''$                   \\
  2012$^{b}$              &    S+C+X      &    B    &    6.0      &    Robust = 0      &    $1.07''\times 0.80''$ PA=-58$^\circ$      \\
  \enddata
  \tablenotetext{a}{Combined data from May 09/30, June 01.} 
  \tablenotetext{b}{Combined data from June 12/16.}
 \end{deluxetable}

  \begin{deluxetable}{lccc}
  \tabletypesize{\normalsize}
  \tablecaption{Upper limits to the Polarization Degree. \label{tbl-gp}}
  \tablewidth{0pt}
  \tablehead{
  \colhead{Knot} & \colhead{S Band (\%)} &  \colhead{C Band (\%)} &  \colhead{X Band (\%)}
  }
  \startdata
  SE       &    $<$ 6   &    $<$ 5    &   $<$ 7     \\
  NW\_C    &    $<$ 6   &    $<$ 9    &   $<$ 12    \\
  NW       &    $<$ 5   &    $<$ 4    &   $<$ 6     \\

  \enddata
  \tablecomments{Upper limits are obtained with the peak intensity of the radio knots and the rms of the maps. A 4-$\sigma$ upper limit was considered.}

 \end{deluxetable}

  \begin{deluxetable}{lccc}
  \tabletypesize{\normalsize}
  \tablecaption{Proper motions of the main knots. \label{tbl-1}}
  \tablewidth{0pt}
  \tablehead{
  \colhead{Source} & \colhead{$\mu$ (mas~yr$^{-1}$)} &  \colhead{{\it v} (km~s$^{-1}$)} &  \colhead{PA~($^{\circ}$)}
  }
  \startdata
  SE\_N       &    152  $\pm$ 1  &    299 $\pm$ 2    &    126 $\pm$ 1 \\
  SE\_S       &    113  $\pm$ 1  &    222 $\pm$ 2    &    132 $\pm$ 1 \\
  NW\_C$^a$   &    280  $\pm$ 10 &    543 $\pm$ 26   &    133 $\pm$ 1 \\
  NW\_C$^b$   &     21  $\pm$ 2  &     42 $\pm$ 3    &    133 $\pm$ 1 \\
  NW          &    104  $\pm$ 1  &    205 $\pm$ 2    &    136 $\pm$ 1 \\

  \enddata

   \tablenotetext{a}{Before 1997.}
   \tablenotetext{b}{After 1997.}
 \end{deluxetable}

\end{document}